\documentclass[10pt,journal,twoclolumn]{IEEEtran}
\usepackage{epsfig}
\usepackage{times}
\usepackage{graphicx,graphics}
\usepackage{url}
\usepackage{amsmath}
\usepackage{amssymb}
\usepackage{amsxtra}
\usepackage{algorithmic}
\usepackage{algorithm}
\usepackage{enumerate}
\usepackage{multirow}
\usepackage{citesort}
\usepackage[small]{caption}
\usepackage[normalem]{ulem}
\usepackage{rotating}
\usepackage{colortbl}
\usepackage{type1cm}
\usepackage{eso-pic}
\usepackage{color}

\makeatletter
\AddToShipoutPicture{%
            \setlength{\@tempdimb}{.5\paperwidth}%
            \setlength{\@tempdimc}{.5\paperheight}%
            \setlength{\unitlength}{1pt}%
            \put(\strip@pt\@tempdimb,\strip@pt\@tempdimc){%
        \makebox(0,0){\rotatebox{45}{\textcolor[gray]{0.75}%
        {\fontsize{1.2cm}{1.2cm}\selectfont{IEEE Trans. Smart Grid 2012 (Accepted)}}}}%
            }%
} \makeatother

\newtheorem{corollary}{\bf Corollary}
\newtheorem{theorem}{\bf Theorem}
\newtheorem{proposition}{\bf Proposition}

\newtheorem{definition}{\bf Definition}

\newcommand{\INDSTATE}[1][1]{\STATE\hspace{#1\algorithmicindent}}

\newcommand{\Rmnum}[1]{\expandafter\@slowromancap\romannumeral #1@}
\makeatother
\newlength{\aligntop}
\newlength{\alignbot}
\makeatletter
\renewenvironment{align}{%
  \vspace{\aligntop}
  \start@align\@ne\st@rredfalse\m@ne
}{%
  \math@cr \black@\totwidth@
  \egroup
  \ifingather@
    \restorealignstate@
    \egroup
    \nonumber
    \ifnum0=`{\fi\iffalse}\fi
  \else
    $$%
  \fi
  \ignorespacesafterend%
  \vspace{\alignbot}\par\noindent
}
\begin{document}
\title{Economics of Electric Vehicle Charging: A Game Theoretic Approach}
\author{Wayes~Tushar,~\IEEEmembership{Student~Member,~IEEE,}
        Walid~Saad,~\IEEEmembership{Member,~IEEE,}
        H.~Vincent~Poor,~\IEEEmembership{Fellow,~IEEE,}
        and~David~B.~Smith,~\IEEEmembership{Member,~IEEE}% <-this % stops a space
\thanks{Wayes Tushar is with the School
of Engineering, Australian National University (ANU), Canberra, ACT
0200, Australia. He is also with National ICT Australia (NICTA).
~e-mail: wayes.tushar@anu.edu.au}% <-this % stops a space
\thanks{Walid Saad is with the Electrical and Computer Engineering
Department, University of Miami, Coral Gables, FL,
USA.~Email: walid@miami.edu}% <-this % stops a space
\thanks{H. Vincent Poor is with
 the School of Engineering and Applied Science, Princeton
University, Princeton, NJ, USA.~Email:~poor@princeton.edu}
\thanks{David Smith is with NICTA, Canberra, ACT, Australia. He also holds an adjunct position at ANU.~Email: david.smith@nicta.com.au.}
\thanks{This work was supported by NICTA. NICTA is funded by the Australian Government as
represented by the Department of Broadband, Communications and the
Digital Economy and the Australian Research Council through the ICT
Centre of Excellence program.}}

\date{}
\maketitle

\begin{abstract}
In this paper, the problem of grid-to-vehicle energy exchange
between a smart grid and plug-in electric vehicle groups~(PEVGs) is
studied using a noncooperative Stackelberg game. In this game, on
the one hand, the smart grid that acts as a leader, needs to decide
on its price so as to optimize its revenue while ensuring the PEVGs'
participation. On the other hand, the PEVGs, which act as followers,
need to decide on their charging strategies so as to optimize a
tradeoff between the benefit from battery charging and the
associated cost. Using variational inequalities, it is shown that
the proposed game possesses a \emph{socially optimal Stackelberg
equilibrium} in which the grid optimizes its price while the PEVGs
choose their equilibrium strategies. A distributed algorithm that
enables the PEVGs and the smart grid to reach this equilibrium is
proposed and assessed by extensive simulations. Further, the model
is extended to a time-varying case that can incorporate and handle
slowly varying environments.
\end{abstract}

%\newpage
 \setcounter{page}{1}
\section{Introduction}
% ------------------------------------------------
\IEEEPARstart{D}{ue} to the growing concerns for energy conservation
and the environment, it is expected that plug-in electric vehicles
(PEVs)\footnote{ PEVs include both battery-only electric vehicles
(BEVs) and plug-in hybrid electric vehicles (PHEVs)~\cite{PB10}.}
will play a major role in the future smart grid (SG) \cite{NR11}.
Therefore, several countries are working on establishing novel PEV
policies and plans because of their significant environmental
advantages and cost savings~\cite{AM10}.

The deployment of PEVs will introduce new challenges in the design
of SGs. These challenges include developing optimal charging
strategies for the connected PEVs, ensuring efficient communications
between PEVs and the grid, and managing energy exchange between
regular loads of the grid and the PEVs. In \cite{NR11}, the optimal
charge control of PEVs is analyzed in deregulated electricity
markets based on a forecast of future electricity prices and the
optimal economic solution for the vehicle owner. In \cite{PS10}, a
real-time pricing algorithm is proposed for smart grids considering
smart meter and energy provider interaction through control messages
exchange. A noncooperative game model for pricing and frequency
regulation in smart grids with electric vehicles is studied in
\cite{HAA00}. Using facility location games, in \cite{FP10}, optimal
locations of battery exchange stations were derived within a
vehicle-to-grid (V2G) network.  An optimization framework is
proposed in \cite{SS11} for enabling the SG to determine the time
and duration of the PEVs charging. In \cite{KT10}, a control
algorithm is developed based on queuing theory to control the
charging of PEVs. {A stochastic programming technique is introduced
in \cite{KEJ2010} to show the impact of charging PHEVs on a
residential distribution grid. The issue of vehicle-to-grid (V2G)
integration and the prospective communication interface to enable
this integration are addressed in \cite{SK10}. In \cite{RSHM2012}, a
mean field game is proposed to investigate the competitive
interaction between electric vehicles in a Cournot market consisting
of electricity transactions to/ from an electricity distribution
network. To manage a large number of PHEVs at a municipal parking
station, an algorithm using particle swarm optimization is proposed
in \cite{WM2011}. An intelligent method for scheduling the usage of
the available storage capacity from PHEVs and electric vehicles is
proposed in \cite{CGK2008}. Other aspects of electric vehicles in
smart grids in terms of energy storage, charging and greenhouse gas
emission reduction are discussed in
\cite{ABB2011,AHEB2010,APBE2009}. Beyond PEVs, many possible demand
side management solutions are also foreseen for the smart
grid~\cite{PD2011} and each of these solutions has a number of
benefits and cost tradeoffs. Subsequently, it is anticipated that
many of these solutions, including the incorporation of electric
vehicles, will co-exist to provide smart energy management for the
power grid.

One of the key challenges of widespread penetration of PEVs in the
power network is the choice of an optimal charging strategy for the
PEVs. This is mainly due to the fact that the integration of the
PEVs into the network has a major impact on the power grid and can
potentially double the average load \cite{PS10,WZ10}. The
simultaneous charging of several PEVs in a particular area can
overload the network and, thus, lead to an interruption of services
for other consumers. The problem of PEV charging and its impact on
the power distribution grid and electricity market have been
addressed in \cite{NR11,PB10,AM10,KT10,SS11} and \cite{PS11}.
However, little has been done to develop distributed models and
algorithms that can capture the interactions between PEVs and the
grid, in a grid-to-vehicle scenario. There is a need to develop
solutions that capture the often conflicting objectives between the
SG, which seeks to maximize its revenue, and the PEVs, which seek to
optimize their charging behavior. Because of the limited grid
capacity and PEVs' energy demands, it is of interest to develop a
model that can capture the decision making process of the PEVs and
the grid when the grid's limited energy needs to be allocated among
the PEVs based on their needs.

The main contribution of this paper is to provide a comprehensive
analytical framework that is suitable for capturing the interactions
between an SG and a number of PEV groups (PEVGs), e.g., parking
lots, which must decide on their charging profiles. We model the
problem as a generalized Stackelberg game in which the SG is the
leader and the PEVGs are the followers. The objective of the PEVGs
is to strategically choose the amount that they need to charge, so
as to optimize a utility that captures the tradeoff between the
charging benefits and the associated costs, given various practical
constraints on the PEVGs and the main grid. Based on the strategy
choices of the PEVGs, the leader aims to optimize its price so as to
maximize its revenues. We analyze the properties of the resulting
game within the studied model, including existence of an equilibrium
and optimality, and show that there exists an efficient (i.e.,
socially optimal) generalized Stackelberg equilibrium. We show that,
due to the coupled capacity constraints between the PEVGs, the
noncooperative followers' game leads to a generalized Nash
equilibrium, and the solution enables the capture of not only the
charging behavior of the vehicles but also the decisions made by the
SG. We propose a novel algorithm that the PEVGs and the grid can
use, in a distributed manner, so as to reach the desired
equilibrium. We also show that the proposed algorithm enables the
system to adapt to time-varying environmental conditions such as
arrival/departure of PEVs. Using extensive simulations, we assess
the properties of the proposed scheme.

The rest of this paper is organized as follows. Section
\ref{system-model} describes the system model. In Section
\ref{NGSE}, we formulate the noncooperative generalized Stackelberg
game and we discuss its properties. In Section \ref{solutionGame},
we propose a distributed algorithm for finding the equilibrium.
Adaptation of the proposed game to time-varying conditions is
discussed in Section~\ref{dynamic-case}.  Numerical results are
analyzed in Section \ref{NumAnalysis} and conclusions are drawn in
Section \ref{conclusion}.
% -------------------------------------------------
\section{system model}\label{system-model}
Consider a power system consisting of a single power grid (i.e., an
SG), several primary and secondary load subscribers or consumers and
a smart energy manager (SEM). Here, the SG refers to the main
electric grid which is connected to the area of interest via one or
more substations. Further, we consider that the SG is servicing a
certain area or groups of primary consumers such as industries,
houses and offices. After meeting the demands of the primary
consumers the grid wishes to sell its excess of energy (if any) to
the secondary users, such as PEVs, connected to it in that area.
Consider a number of groups of PEVs (hereinafter, we use the term
PEVG to denote a group of PEVs, acting as a single PEV entity),
which are connected to the grid at peak hours of energy demand e.g.,
from {$12$ pm} to $4$ pm\footnote{The proposed scheme is not only
restricted to the considered period but can also apply for any time
duration.} \cite{EC08}. {The total charging period for the PEVGs is
considered to be divided into multiple time slots. Each time slot
has a duration of anywhere between $5$ minutes and half an hour
based on the changing traffic conditions of the PEVs in the
group~\cite{WH2001}}. For a particular time slot, the power grid has
a maximum energy $C$ that it can sell to the connected ${N}$ PEVGs,
allowing them to meet their demand\footnote{While this paper is
focused on the interactions between PEVGs and the grid, when the
grid has energy that can be provided to the PEVGs, it can also be
extended to cases in which there are multiple energy sources beyond
the main grid.}. The power grid will set an appropriate price $p$
(per unit of energy) for selling its surplus so as to optimize its
revenue.

Each PEVG $n\in\mathcal{N}$, where $\mathcal{N}$ is the set of all
$N$ PEVGs,  will request a certain amount of energy $x_n$ from the
grid so as to meet its energy requirements (e.g., to go back home
after office work). This demand of energy may vary for the PEVs in
the group based on different parameters such as the battery capacity
$b_n$ of the PEVG, the available energy in the PEVG's battery at the
time of plug-in to the grid, the price $p$  per unit of electricity
and the nature of usage (e.g., two identical PEVGs with different
travel plans may need different amounts of energy). During peak
hours, the available energy for servicing PEVGs is often limited
\cite{EC08}, and, thus, the PEVGs will request only the amount
needed due to their immediate need for charging. Since the  net
energy $C$ available for the PEVGs at the grid is fixed, the demands
of the PEVGs must satisfy
\begin{align}
\sum_n x_n \leq C. \label{DemandCond}
\end{align}

Given the amount requested by the PEVGs, the SG sets a price
$p$ per unit of energy so as to optimize its revenue from selling
energy by strategically choosing its price per unit of energy.
Although the SG can choose the price $p$ within any range to
maximize its total revenue, a very large $p$ may compel a PEVG to
withdraw its demand from the grid and to search for alternate
markets or wait until the prices drop. Therefore, an optimal
price needs to be chosen by the grid operator which is not very
high (to avoid losing customers) and not very low (to avoid losing
revenue), in an effort to maximize its profit.

To successfully complete the energy trading, the PEVGs and the grid
interact with each other and agree on the energy exchange
parameters, such as the selling price and the amount of energy
demanded, that meet the objectives of the both sides. The amount of
requested energy $x_n$ is determined by the physical characteristics
of PEVG $n$ as well as by the tradeoff between the potential benefit
that PEVG $n$ is expecting from buying $x_n$ and the selling price
$p$ of the grid. Moreover, the selling price $p$ per unit of energy
is strongly dependent upon the amount demanded by the PEVGs as well
as the number of PEVGs that are connected to the grid. This is due
to the fact that the amount of energy available from the grid is
fixed and, thus, as the number of PEVGs increases, the amount that
each PEVG can acquire becomes smaller. As a result, the grid can set
a higher price to increase its revenues.

It is clear that the demands of the connected PEVGs are coupled
through the energy constraint in \eqref{DemandCond}, and are also
dependent on the physical characteristics and capabilities of the
PEVGs. Also it is important to note that the price set by the grid
is dependent on the demand of each PEVG. Thus, the main challenges
faced when developing an approach that can successfully capture the
decision making process of both the PEVGs and the grid are: i)-
modeling the decision making processes and the interactions between
the connected PEVGs in the network given the constraint in
\eqref{DemandCond}; ii)- developing an algorithm that enables the
PEVGs in the network to strategically decide on the amount of energy
that they will request from the grid so as to optimize their
satisfaction levels given the constraint in  \eqref{DemandCond}; and
iii)- enabling the grid to optimize its price while capturing the
tradeoff between the PEVGs' participation and revenue maximization.
To address these challenges we propose a framework based on
noncooperative game theory.
% ---------------------------------------
\section{Noncooperative generalized Stackelberg game}\label{NGSE}
% ---------------------------------------
\subsection{Game formulation}\label{game_formulation}
\label{sub-section_game} To formally study the interactions between
the SG and the PEVGs, we use a Stackelberg game \cite{BA00} which is
a type of noncooperative game that deals with the multi-level
decision making process of a number of independent decision makers
or players (followers) in response to the decision taken by a
leading player (the leader)~\cite{BA00}. Hence, we formulate a
noncooperative Stackelberg game in which the SG is the leader and
the PEVGs are the followers. This game is defined by its strategic
form,
 $\Gamma = \left\lbrace (\mathcal{N}\cup\{\text{SG}\}), \left\lbrace X_n \right\rbrace_{n\in\mathcal{N}},\left\lbrace U_n\right\rbrace_{n\in\mathcal{N}},L(p),p\right\rbrace$, having the following components:
 \begin{enumerate}[(i)]
 \item The PEVGs in $\mathcal{N}$ act as the followers in the game and respond to the price set by the SG.
 \item The strategy of each PEVG $n\in\mathcal{N}$ which corresponds to the amount of energy demanded $x_n\in\mathbf{X_n}$ from the grid satisfying the constraint $\sum_{n}x_n\leq C$.
 \item The \textit{utility function} $U_n$ of each PEVG $n$ that captures the benefit of consuming the demanded energy $x_n$.
 \item The \textit{utility function} $L(p)$ for the SG (leader of the game), which captures the total profit that the grid can receive by selling the surplus energy with price $p$.
 \item The price $p$ per unit of energy charged by the SG.
 \end{enumerate}
 % --------------------------------------------------------------
\subsubsection{Utility Function of a PEVG}\label{utility of PEVs-property}
For each PEVG $n\in\mathcal{N}$, we define a utility function $U_n(
x_n, \mathbf{x_{-n}}, s_n, b_n, p)$, which represents the level of
satisfaction that a PEVG obtains as a function of the energy it
consumes. Here $b_n$ is the battery capacity of PEVG $n$ and $x_n$
is the requested energy from the grid; $s_n$ is the
satisfaction parameter
 of PEVG $n$, which is a measure of the satisfaction the PEVG can achieve from consuming one unit of energy. This $s_n$ may
depend on PEVG's battery state at the time of plug-in to the grid,
the available energy at the grid and/or the travel plan of the
corresponding PEVG. For example, a PEVG $1$ having less need for
energy than another PEVG $2$ (e.g., due to having a fuller battery)
will need less energy than PEVG $2$ to attain the same satisfaction
level (i.e., $s_2 < s_1$). The energy demand may vary based on the
battery capacity and/or the satisfaction parameter of each PEVG. The
price per unit of energy can also affect the demand of the PEVG.
Thus, the properties that the utility of a PEVG must satisfy are as
follows:
\begin{enumerate}[(i)]
\item \label{p-1}The utility functions of the PEVGs are considered to be non-decreasing because each PEVG is interested in consuming more energy if
possible unless it reaches its maximum consumption level.
Mathematically,
\begin{align}
\frac{\delta U_n( x_n, \mathbf{x_{-n}}, s_n, b_n, p)}{\delta x_n}\geq 0.
\label{property-1}
\end{align}
\item  \label{p-2} The marginal benefit of a PEVG is considered as a
non-increasing function, as the level of satisfaction of the PEVGs
gradually gets saturated as more energy is consumed, i.e.,
\begin{align}
\frac{\delta^2 U_n( x_n, \mathbf{x_{-n}}, s_n, b_n, p)}{\delta {x_n}^2}\leq 0.
\label{property-2}
\end{align}
\item \label{p-3} Hereinafter, we consider that, for a fixed consumption level $x_n$, a
larger $b_n$ implies a larger $U_n(x_n,\mathbf{x_{-n}}, s_n, b_n, p)$ and a
larger $s_n$ leads to a smaller $U_n( x_n, \mathbf{x_{-n}}, s_n, b_n, p)$. So
we have
\begin{align}
\frac{\delta U_n(x_n, \mathbf{x_{-n}}, s_n, b_n, p)}{\delta b_n}> 0,
\label{property-3a}
\end{align}
and
\begin{align}
\frac{\delta U_n( x_n, \mathbf{x_{-n}}, s_n, b_n, p)}{\delta s_n}<0.
\label{property-3b}
\end{align}
\item \label{p-4}The price per unit of energy set by the grid affects
the utilities of the PEVGs and the utility of a PEVG decreases with
a higher price. That is,
\begin{align}
\frac{\delta U_n( x_n,\mathbf{x_{-n}}, s_n, b_n, p)}{\delta p}<0.
\label{property-4}
\end{align}
\end{enumerate}
In this work we consider the particular utility
\begin{align}
U_n(x_n, \mathbf{x}_{-n}, s_n, b_n, p) = b_nx_n - \frac{1}{2}s_n{x_n}^2 -
px_n, \label{utility_function}
\end{align}
% -----------------------------------------------------
{where $x_n\in[0,C-\sum_{k=1,k\neq n}^N x_k]$ and
$\mathbf{x_{-n}}=[x_1, x_2,...,x_{n-1}, x_{n+1},...,x_N]$}, although
many of our results can be generalized.

From \eqref{utility_function}, the utility of  PEVG $n$ is affected
by its battery capacity $b_n$. This is due to the fact that a PEVG
with higher $b_n$ will have higher marginal utility and thus, needs
to consume more energy to reach its maximum satisfaction level
\cite{MF99}. The utility also depends on the satisfaction parameter
$s_n$ of the PEVG. PEVGs with the same capacity but with a different
satisfaction parameter will have different marginal utilities and,
thus, will be satisfied by different amounts of energy. To this
end, we assume that a PEVG does not consume any energy beyond its
maximum satisfaction level, i.e., $U_n = 0$ if $x_n>({x_n}^* -
x_{n}^{\text{ini}})$, where $x_{n}^{\text{ini}}$ is the initial
energy in PEVG's battery at the time of plug-in to the grid and
${x_n}^*$ is the energy that maximizes its utility within the given
constraint in \eqref{DemandCond}.

\subsubsection{Utility Function of the Power grid}
A PEVG $n$ that consumes $x_n$ MWh of electricity during a
designated period of time at a rate $p$ per MWh is charged $px_n$
which is the cost imposed by the SG on the PEVG. The objective of
the SG is to maximize its revenue by selling the available energy
surplus to the PEVGs after meeting the demand of its primary
consumers, and also to control the nature of energy consumption of
the PEVGs. While this energy surplus $C$ is fixed, the SG wants to
set a price $p$ per unit of energy so as to optimize its
revenue, given the demands of the PEVGs. Thus, we assume the utility
function for the grid is
\begin{align}
L(p,x_n(p)) = p\sum_n x_n, \label{utility_grid}
\end{align}
  which
captures the total revenue of the grid when selling the energy
required by all PEVGs at a price $p$ per unit of
energy.

In this proposed game, the SG can control the price $p$ per unit of
energy it wishes to sell. Connected PEVGs respond to the price by
demanding a certain amount of energy, given the constraint in
\eqref{DemandCond}, so as to maximize their utilities. Thus, for a
fixed price $p$, the objective of any PEVG $n$ is
\begin{align}
\displaystyle\max_{x_n\in(C-\mathbf{x_{-n}})} U_n(x_n,
\mathbf{x_{-n}}, s_n, b_n, p), \nonumber \\\text{s. t.}~\displaystyle\sum_n
x_n\leq C.
\end{align}
Here, we can see that the amount of energy demanded by each PEVG $n$
depends, not only on its own strategies and price but also on the
demand of other PEVGs in the network through \eqref{DemandCond} and
the constraint is the same for all players. This leads connected
PEVGs to engage in a noncooperative resource sharing game, which is
a jointly convex generalized Nash equilibrium problem (GNEP)  due to
the same shared constraint \eqref{DemandCond}. Note that, in game
theory, a noncooperative game in which the players' actions are
coupled solely through the constraints, such as in the proposed
model, is a special class of games whose solution is the generalized
Nash equilibrium \cite{FF07,DA11}, and hence, the proposed
followers' game, for any price $p$, is a noncooperative resource
sharing game whose solution is the generalized Nash equilibrium
(GNE). Then, given all the PEVGs' demands are at the GNE, the
leader, i.e., the grid, chooses the price to maximize its revenue.
Thus, for the given GNE demands of the PEVGs, the objective of the
grid is
\begin{align}
\displaystyle\max_{p}L(p) = \displaystyle\max_{p}
\displaystyle\sum_n px_n.
\end{align}
Thus, one suitable solution for the formulated game
$\Gamma$ is the Stackelberg equilibrium at which the leader reaches
its optimal price, given the followers' optimal response at their
GNE. At this equilibrium, no player (leader or follower) can improve
its utility by \emph{unilaterally} changing its strategy. In
classical Stackelberg games, the followers typically choose their
Nash equilibrium strategies. In our model, due to the coupled
strategies as per \eqref{DemandCond}, the PEVGs need to seek a GNE
instead of a classical Nash equilibrium. To this end, hereinafter,
we refer to our game as a generalized Stackelberg game (GSG) whose
solution is the generalized Stackelberg equilibrium (GSE) in which
the followers reach a GNE.

\definition\label{definitionSoln1}Consider the GSG $\Gamma = \left\lbrace (\mathcal{N}\cup\textrm{SG}),
\left\lbrace X_n \right\rbrace_{n\in\mathcal{N}},\left\lbrace
U_n\right\rbrace_{n\in\mathcal{N}},L(p),p\right\rbrace$ defined in
\ref{game_formulation} where $U_n$ and  $L(p)$ are given by
\eqref{utility_function} and \eqref{utility_grid} respectively. A
set of strategies ($\mathbf{x^*},p^*$) constitutes the GSE of this
game, if and only if it satisfies the following set of inequalities:
\begin{align}
U_n({x_n}^*,\mathbf{{x_{-n}}^*}, s_n, b_n, p^*)\geq
U_n({x_n},\mathbf{{x_{-n}}^*}, s_n, b_n, p^*),\nonumber\\\forall
{x_n}^*\in\mathbf{x^*},\hspace{0.2cm} n\in \mathcal{N}, \sum_n
x_n\leq C \label{eqn_soln1}
\end{align}
and
\begin{align}
L(p^*,\mathbf{x^*})\geq L(p,\mathbf{x^*}). \label{eqn_soln2}
\end{align}
Thus, when all the PEVGs' demands are at the GSE, no PEVG can
improve its utility by deviating from its GSE demand and similarly,
no price other than the optimal price\footnote{{The optimal price
$p^*$ maximizes the utility of the SG for the given GNE demand
vector $\mathbf{x^*}$ of the PEVGs in the smart grid network.}}
$p^*$ set by the grid at the GSE, can improve the utility for the
grid.
% -------------------------------------------------------------
\subsection{Existence and efficiency of GSE}
\label{existenceUniqueness} In noncooperative games, the existence
of an equilibrium solution (in pure strategies) is not always
guaranteed \cite{BA00}. Therefore, for our follower game, we need to
investigate the existence of the GNE in response to a price $p$.
Specifically, we are interested in investigating the existence and
properties of a variational equilibrium (VE), which is a type of GNE
to be defined below (see also \cite{FF07}), for our case. This is
due to the fact that a VE is more socially stable than another GNE
(if there exists any) and thus, is a desirable target for any
algorithm to achieve \cite{DA11}. Particularly for the proposed
case, where a number of PEVGs in the SG network are demanding energy
from a constrained reserve, an efficient VE would be the most
appropriate solution to be considered. Hereinafter, we use VE and
GNE interchangeably.

\theorem\label{theorem2} For a fixed price $p$, a socially optimal
VE exists in the proposed game $\Gamma$ between the PEVGs connected
to the grid.

\begin{IEEEproof}
First, clearly, by adding the quantity $\sum_{m\neq n}(b_m
x_m-\frac{1}{2}s_m{x_m}^2)-\sum_{m\neq n}px_m$ to $U_n$ in
\eqref{utility_function} and treating the resulting utility function
as the new objective function for PEVG $n$ will not affect the
solution \cite{TBINFO02}. Thus, the original game is equivalent to
one in which all PEVGs have the same utility function,
\begin{eqnarray}
U(x_1,...,x_{{N}}; s_1, ..., s_N; b_1, ..., b_N;
p)\nonumber\\=\sum_{m=1}^{{N}}\left(b_mx_m-\frac{1}{2}s_m{x_m}^2\right)-p\sum_{m=1}^{{N}}x_m.
\label{genral_utility}
\end{eqnarray}
{Hence, to determine the socially stable outcome of the game, the
existence of a solution that maximizes \eqref{genral_utility} is our
main concern.}

Using the method of Lagrange multipliers \cite{DB95}, the
Karush-Kuhn-Tucker (KKT) conditions for the $n^{\text{th}}$ player
GNEP is given by

\begin{align}
-{\nabla_{x_{n}}}U_n(x_n,\mathbf{x}_{-n}, s_n, b_n,
p)+{\nabla_{x_{n}}}\left(\displaystyle\sum_nx_n-C\right)\lambda_n=0,\nonumber\\
~\lambda_n\left(\displaystyle\sum_nx_n-C\right)=0,\lambda_n\geq
0\label{KKT_GNEP}
\end{align}
where $\lambda_n$ is the Lagrange multiplier for PEVG $n$.

First, we note that, for a fixed price $p$, the followers' game
admits a jointly convex GNEP, hence, the solution of the GNEP with
\eqref{DemandCond} can be found via a variational inequality
VI$(\mathbf{X,F})$. This essentially reduces to determining a vector
$z^*\in\mathbf{X}\subset\mathbb{R}^n$, such that $\langle
\mathbf{F(z^*),z-z^*}\rangle\geq 0$, for all $z\in\mathbf{X}$ where
$\mathbf{X}$ is the set in the definition of joint convexity and
$\mathbf{F(x)}=-(\nabla_xU_n(\mathbf{x}))_{n=1}^{{N}}$ \cite{DA11}.
The solution of VI$(\mathbf{X,F})$ is a variational equilibrium.

Now the KKT conditions {can be written as~\cite{FF07}}
\begin{align}
\mathbf{F(x)} + \lambda\nabla_x(\sum_n x_n - C) =0,\nonumber\\
\lambda(\sum_n x_n - C)=0,~\lambda\geq 0. \label{kkt_VI}
\end{align}
Note that the subscript on {Lagrange multiplier} $\lambda$ is
dropped in \eqref{kkt_VI} . This is due to the fact that the
solution of a jointly convex {GNEP} is a VE if and only if the
shared constraint has the same multiplier $\lambda$ for all players
\cite{FF07}.

Now from the definition of $\mathbf{F}$ \cite{DA11}, we have
\begin{align}
\mathbf{F} = \left[ \begin{array}{c}
s_1x_1 +p - b_1\\
s_2x_2 + p - b_2 \\
\vdots \\
s_nx_n + p - b_n
\end{array} \right].
\label{defF}
\end{align}

Therefore, the Jacobian of $\mathbf{F}$ is
\begin{align}
\mathbf{JF}=
\begin{bmatrix}
s_1 & 0 & ..... & 0\\
. & s_2 & ..... & .\\
. & . & ..... & .\\
. & . & ..... & .\\
0 & 0 & .....& s_n
\end{bmatrix}.
\label{JF}
\end{align}
{$\mathbf{JF}$ is a diagonal matrix with all positive diagonal
elements.} Hence, $\mathbf{JF}$ is positive definite on
$\mathbf{X}$, and so, $\mathbf{F}$ is strictly monotone. Thus, {the
GNEP} admits a unique global VE solution \cite{FF07}.

Because of the jointly convex nature of the {GNEP} the VE is the
unique global maximizer of \eqref{genral_utility} \cite{FF07}, which completes the proof.
\end{IEEEproof}

As a result, from Theorem~\ref{theorem2}, the GSE, in which the SG
sets its optimal price in response to the VE demands of the PEVGs,
admits the socially optimal solution of the proposed game.

\section{{Proposed solution and algorithm}}\label{solutionGame}
In this section, we formulate the GNEP among the followers as a
variational inequality (VI) problem\footnote{Given
$\mathbf{X}\subseteq \mathbb{R} ^n$ and
$\mathbf{F}:\mathbb{R}^n\rightarrow\mathbb{R}^n$, the
VI$(\mathbf{X,F})$ consists of finding a vector $z^* \in \mathbf{X}$
such that $\langle \mathbf{F(z^*),z-z^*}\rangle\geq 0$, for all
$z\in\mathbf{X}$.} and propose an algorithm  that leads to the
socially optimal VE. Note that the VE further leads to the GSE state
of the game as defined in Definition~\ref{definitionSoln1}. Now, we
first state the following corollary and then explain the solution
method for the considered GNEP.

\corollary\label{prop-2} The VI associated with the proposed GNEP
of the connected PEVGs for a price $p$ is a strongly monotone VI and
thus, the unique VE can be calculated by solving a monotone VI.
\begin{IEEEproof}
{By Theorem~\ref{theorem2}, we know that the VI associated with the
proposed GNEP of the connected PEVGs for any fixed price $p$ is a
strongly monotone VI and the VE is unique. It is shown in
\cite{FF07} that the solution of a VE can be calculated by solving a
monotone VI. Hence, the unique VE solution of the PEVGs' GNEP of
energy demand within constraint \eqref{DemandCond} can be calculated
by solving the strongly monotone VI($\mathbf{X,F}$).}
\end{IEEEproof}

For solving the monotone VI in our proposed game, we consider the
Solodov and Svaiter (S-S) hyperplane projection method \cite{MV99,FT03}. In the
S-S method, two projections per iteration are required using a
geometric interpretation (see \cite{MV99}).  This hyperplane
projection algorithm works as follows \cite{MV99}: Suppose we have
$x^k$, which is a current approximation to the solution of
VI$(X,F)$. First the point $\text{Proj}_X[x^k - F(x^k)]$ is
computed\footnote{$\text{Proj}_X(z) = \arg\min\{||w-z||,~w\in
\mathbf{X}\}~\forall z\in\mathbb{R}^n.$}. Next, the line segment
between $x^k$ and $\text{Proj}_X [x^k - F(x^k)]$ is searched for a
point $z^k$ such that the hyperplane $\partial H_k := \lbrace
x\in\mathbb{R}^n | \langle F(z^k), x - z^k \rangle = 0\rbrace$
strictly separates $x^k$ from any solution $x^*$ of the problem. A
computationally inexpensive Armijo-type procedure \cite{AJ66} is
used in this S-S algorithm to find such $z^k$. Once the hyperplane
is constructed, the next iterate $x^{k+1}$ is computed by projecting
$x^k$ onto the intersection of the feasible set $X$ with the
hyperspace $H_k := \lbrace x\in\mathbb{R}^n : \langle F(z^k), x -
z^k \rangle \leq 0 \rbrace$,~i.e. $X\cap H_k$, which contains the
solution set \cite{MV99} \cite{FT03}.

Next, we show how the PEVGs reach the VE, for a price $p$, following
the optimization of price by the grid when all PEVGs are in VE.
Then, we detail the algorithm at the end of this section.
% -------------------------------------------------
\subsection{GNE for a fixed $p$}
From \eqref{genral_utility} and \eqref{kkt_VI}, for any PEVG $n$,
the solution of the KKT system of variational inequalities is
\begin{align}
b_n - s_nx_n - p -\lambda = 0, \label{eqn_kktVI}
\end{align}
where
\begin{align}
\lambda\left(\displaystyle\sum_n x_n -C\right)=0;~\lambda\geq0.
\label{const_eqn}
\end{align}
For $\lambda>0$, the inequality constraint in \eqref{const_eqn}
becomes as equality and hence at the VE,
\begin{align}
\displaystyle\sum_n x_n = C. \label{equality_const}
\end{align}

Thus, for a fixed $p$ at the grid, the sum of demands of all
the PEVGs connected to the grid at the VE is equal to the total
energy $C$ available at the grid.

At the peak hour of demand, energy in the grid is a scarce commodity
and, hence, all the PEVGs compete with one another for a fair
allocation of the available grid energy. Thus, for the formulation
of the proposed game between the PEVGs the available energy at the
grid should be less than the total energy consumption capacity of
the connected PEVGs. This is essential for avoiding the trivial case
in which all the PEVGs should get an allocation equal to their
capacity. From \eqref{eqn_kktVI}, we have
\begin{align}
b_n-s_nx_n-p>0\nonumber\\
\text{i.e.,}~ b_n>s_nx_n+p. \label{condition_capacity}
\end{align}
Taking all ${N}$ PEVGs connected to the grid into consideration,
\eqref{condition_capacity} becomes
\begin{align}
\displaystyle\sum_n b_n > p{N}+\displaystyle\sum_n s_n x_n,
\label{condition_capacity_total}
\end{align}
which leads to the following proposition:

\proposition\label{proposition-soln2} To achieve the maximum
utilities at the VE, within the constraint in (1), the total
capacities of the $N$ grid connected PEVGs must be greater than
their total VE demand plus a constant equal to $pN$.

For the special case in which the PEVGs have different capacities
 (i.e., $b_n$ is different for each $n$) but the same satisfaction
parameter (i.e., $s_n = s$ for all $n\in\mathcal{N}$),
\eqref{condition_capacity_total} becomes
\begin{eqnarray}
\displaystyle\sum_n b_n>p{N}+s C\\
\text{where}~\displaystyle\sum_n x_n = C,~\text{from}~
\eqref{equality_const}\nonumber. \label{condition_capacity_samePEV}
\end{eqnarray}

Now, while Proposition \ref{proposition-soln2} holds for a price $p$
at the grid, from \eqref{eqn_kktVI}, the demand of the PEVGs at the
VE is given by
\begin{align}
{x_n}^* (p) = \frac{b_n-(p+\lambda)}{s_n}, \label{demand_at_GNE}
\end{align}
 where
\begin{align}
\lambda = b_n-s_n{x_n}^*-p~\text{for any}~ n\in\mathcal{N}.
\label{lambda-gne}
\end{align}
% ---------------------------------------------------------------

\subsection{Price optimization}
Having analyzed the followers' game, we now show how the SG can set
its optimal price $p^*$ given the VE of the PEVGs.

For the KKT system of VIs described in \eqref{eqn_kktVI} and
\eqref{const_eqn}, the selling price for per unit energy is
\begin{align}
p\leq b_n - s_nx_n. \label{p-for-GNG}
\end{align}
Again, from \eqref{demand_at_GNE}, the demand for energy by PEVG $n$
at the VE is ${x_n}^*$. So the price per unit of energy satisfies
\begin{align}
p\leq b_n-s_n{x_n}^*. \label{p-for-GNE}
\end{align}
Now, with the condition in \eqref{p-for-GNE} and the utility of the
SG from \eqref{utility_grid}, which is $L(p) =
p\displaystyle\sum_nx_n$ over $p\geq 0$, this dictates that the
revenue-maximizing price of the grid should be the upper limit of
\eqref{p-for-GNE}. Thus, the optimal price\footnote{{From
\eqref{lambda-gne}, at $p=p^*$, the slack variable $\lambda=0$.
Hence $\lambda_n$ converges to $\lambda=0$,$~\forall n$ as the game
reaches its GSE.}} of the proposed GSG is
\begin{align}
p^* = b_n-s_n{x_n}^*. \label{GSNE_p}
\end{align}
%% ----------------------------------------------
\subsection{Proposed algorithm}
% ----------------------------------------------
\begin{algorithm}[t]
\caption{Algorithm to reach GSE}
\begin{algorithmic}
\scriptsize \STATE \textbf{\emph{1. Solving VI}} \STATE Each PEVG
$n\in\mathcal{N}$ submits its initial demand ${x_n}^{\text{ini}}$ to
the SEM. \STATE \textbf{repeat} \INDSTATE a) The SEM checks
${\lambda_n}^k$ for the demand ${x_n}^k$ of all\\\hspace{0.6cm}
$n\in\mathcal{N}$ using \eqref{kkt_VI}.\INDSTATE b) Each PEVG
$n\in\mathcal{N}$ updates its demand ${x_n}^k$ using the
\\\hspace{0.6cm}S-S hyperplane projection method \cite{FT03}. \INDSTATE
$\mathrm{S-S~ method}$\INDSTATE[2]i) The PEVG $n$ computes the
projection $r({x_n}^k)$.\INDSTATE[2]ii) The PEVG $n$ updates
${x_n}^{k+1}$ equal to $r({x_n}^k)$\\\hspace{1cm} if $r({x_n}^k) =
0$ and submit to the SEM. \INDSTATE
$\mathrm{otherwise}$\INDSTATE[2]iii) The PEVG $n$ determine the
hyperplane ${z_n}^k$ and \\\hspace{1cm}the half space ${H_n}^k$ from
the projection. \INDSTATE[2]iv) The PEVG $n$ updates its demand
${x_n}^{k+1}$ from \\\hspace{1cm}the projection of its previous
demand ${x_n}^k$ on to \\\hspace{1cm}$X\cap {H_n}^k$ and submit to
the SEM. \STATE \textbf{until} all $\lambda_n$ converge to
$\lambda\geq 0$.

\STATE\emph{The SEM determines the VE demand of the PEVGs.}

\STATE\textbf{\emph{2. Optimizing Price}}\INDSTATE a) The SEM
submits the VE demand of the PEVGs \\\hspace{0.5cm}to the
grid.\INDSTATE b) The grid optimizes its price $p$ to $p^*$ using
\eqref{GSNE_p}.

\STATE\emph{The VE demand and price of the GSG are achieved.}

\end{algorithmic}
\label{algorithm1}
\end{algorithm}
% ----------------------------------------------
In order to reach the equilibrium, the PEVGs and the smart grid must
make their strategy choices with little communication between one
another. To this end, we propose an algorithm that all the PEVGs and
the grid can implement in a distributed fashion and reach the
efficient GSE of the game. We note that, in a jointly convex GNEP
where $\mathbf{F(x)}=-(\nabla_xU_n(\mathbf{x}))_{n=1}^{{N}}$ is
strongly monotone, such as our proposed game, the solution of the VI
converges to a unique VE \cite{FF07} when the demand of each PEVG
$n$ is such that the parameter $\lambda_n$ in \eqref{kkt_VI} for all
$n\in\mathcal{N}$ possesses the same value $\lambda\geq 0$. In other
words, if the parameter $\lambda_n$ converges to a single value
$\lambda\geq 0$ for all $n\in\mathcal{N}$, then $\mathbf{x^*}$, the
demand vector of the PEVGs contains the VE demand of the PEVGs.

For our game, we can use this property and propose an algorithm in
which each PEVG updates its demand iteratively, until all
$\lambda_n$ converge to a single value $\lambda\geq 0$. In this
algorithm we use the hyperplane projection method to solve the
proposed VI problem. By using this technique, we guarantee that our
algorithm always converges to a non-empty solution if $\mathbf{F}$
is strongly monotone \cite{AN08} which is always verified in our
game, as previously shown. Thus, the proposed algorithm is
guaranteed to converge to a unique solution of the game, given the
demand constraints of the PEVGs and the grid's capacity $C$. As we
explained in Section \ref{NGSE}, this convergence implies that the
proposed GSG reaches its GSE as soon as the grid optimizes the price
using \eqref{GSNE_p} for the given VE demand of the PEVGs.

The proposed algorithm uses the S-S hyperplane projection method
\cite{FT03} to calculate the demand at the VE for a price $p$. Each
PEVG and the SG can implement the proposed algorithm to reach the
GSE in a distributed fashion with the assumption that the SEM can
communicate with both the grid and the PEVGs. The SEM can use any
vehicle to grid infrastructure technique, \cite{FP10}, for this
communication. As soon as any PEVG $n$ is connected to the grid, the
SEM receives the utility parameters $b_n$ and $s_n$ using V2G. The
algorithm starts with the announcement of the available energy $C$
and the price per unit energy $p$ by the grid. At any given
iteration $k$, in response to this price $p$, each PEVG $n$ updates
its demand for a particular amount of energy ${x_n}^k$ from the
fixed amount $C$ of the grid using the S-S method. The SEM gets the
price $p$ from the grid and checks the parameter ${\lambda_n}^k$
using \eqref{kkt_VI}. To enable the SEM to check the value of
${\lambda_n}^k$, each PEVG $n$ submits its demand ${x_n}^k$ to the
SEM at the end of iteration $k$. The process continues until all the
PEVGs make their demands such that $\lambda_n = \lambda\geq
0~\forall n$. The demand of the PEVGs reaches a VE as the SEM
determines that $\lambda_n = \lambda\geq 0$ for all
$n\in\mathcal{N}$. Then, the SEM submits the VE demand of the PEVGs
to the grid and the grid sets the optimal price $p^*$ using
\eqref{GSNE_p} and, thus, the proposed GSG reaches the GSE.

After the execution of the algorithm the demand of each PEVG $n$
reaches its equilibrium value ${x_n^*}$ which is given by
\begin{align}
{x_n}^*(p^*) = \frac{b_n - p^*}{s_n}, \label{demand_GSNE}
\end{align}
with the optimal price $p^*$. This is the equilibrium of the game.
\section{{Adaptation to Time-Varying Conditions}}\label{dynamic-case}
Here, we extend our approach so as to accommodate time-varying
conditions using a discrete time feedback Stackelberg game
model with dependent followers~\cite{PLM2004}. We assume that the
number of vehicles at a given location, e.g., a parking lot, changes
gradually in real time with a moderate time duration (for example,
$5$ minutes to $30$ minutes)~\cite{WH2001}. We also assume that the
available energy from the grid varies across moderate time
intervals, e.g., once in an hour~\cite{OA2010}. Hence, using a
discrete time feedback Stackelberg game, we can capture changes
in system variables from one time instant to another. Prior to
discussing how the proposed scheme can be adapted in a
time-varying environment, we define the following parameters:
\\$C_t:$ the state variable of the game, which indicates the state of
available charge at time instant $t$.\\ $b_n^t:$ the battery
capacity of PEVG $n$, which depends on the aggregate quantity of the
number of PEVs in the group at instant $t$.\\ $s_n^t:$ the
satisfaction parameter\footnote{Here, we consider that the
satisfaction parameter changes randomly between consecutive time
slots, for example, due to the random change of vehicles in a
parking lot from one time slot to the next.} of
PEVG $n$ at $t$.\\ $x_n^t:$ the energy demand by PEVG $n$ at $t$.\\
$p_t:$ the price per unit of energy at instant $t$.\\ $\mathbf{x^t}
= (x_1^t, x_2^t,...,x_N^t):$ the
vector of demands of all the PEVGs in the network at time $t$.\\
$\mathbf{x_{-n}^t} = (x_1^t,...,x_{n-1}^t, x_{n+1}^t, ...,x_N^t):$
the vector of strategies of all PEVGs except PEVG $n$ at instant
$t$.\\ $L_{T-t} = \sum_t p^t\sum_n x_n^t:$ the payoff function of
the grid which it wants to maximize over the entire peak hour
duration.\\ $U_{T-t}
=\sum_t\sum_n\left(b_n^tx_n^t-\frac{1}{2}s_n^t{x_n^t}^2-p_tx_n^t\right):$
the joint utility function of the PEVGs in the network.}

Consequently, the state transition equation for the time-varying
system can be defined as~\cite{KB2011}
\begin{eqnarray}
C_{t+1}=f_t(C_t,p_t,\mathbf{x^t}),~t=0,1,2,...,T-1,\label{state-equation}
\end{eqnarray}
where $t$ is an integer time index, and $T$ is the entire peak hour
duration. For a feedback Stackelberg game, $C_t$ is the information
about the available energy supply at time $t$, which is gained by
the grid from \eqref{state-equation} and fed back to the PEVGs
through the SEM~\cite{KB2011}. The state of available energy supply
at any instant $t$ is a function of the demand for energy by the
PEVGs, and the energy available at the previous time slot. The state
transition from $t$ to $t+1$ takes place when a change occurs either
in the number of PEVs in the PEVG, or in the available energy from
the grid. The objective of both the SG and the PEVGs is to choose
their strategies so as to maximize their utilities in each time
interval, and thus to maximize their total payoffs in the entire
time horizon of peak hours. Hence, the problems of payoff
maximization of the players for a discrete time feedback Stackelberg
game can be formally expressed as
\begin{eqnarray}
\max L_{T-t} = \sum_t \max_{p_t}p_t\sum_nx_t^n,\label{dyn-grid}
\end{eqnarray}
for the SG, and
\begin{eqnarray}
\max U_{T-t} =\sum_t \max_{x_t^n}
\sum_n\left(b_n^tx_n^t-\frac{1}{2}s_n^t{x_n^t}^2-p_tx_n^t\right)\label{dyn-phevg}\\~\sum_nx_n^t\leq
C_t~\text{at}~t=0,1,2,...,T,\nonumber
\end{eqnarray}
for the PEVG $n$. The objective functions in \eqref{dyn-grid} and
\eqref{dyn-phevg} refer to a feedback Stackelberg game with Nash
game constraints in the lower level decision making
process~\cite{PLM2004} (similar to the game proposed in
Section~\ref{NGSE} for a single time instant) over the whole time
horizon. Here, we assume that the leader of the game can perfectly
gain the information about $C_t$ from
\eqref{state-equation}~\cite{BT1983}. At any instant $t$, the SG
gets the information about the volume of the parking lot (i.e.,
$b_n^{t-1}$ and $s_n^{t-1}$ for all $n= 1,2,..,N$), the demand of
the PEVGs in the previous time slot $t-1$ as well as the per unit
price $p_{t-1}$. Then, it estimates the amount of energy that needs
to be provided to the PEVGs at $t$ through \eqref{state-equation}.
The SG feeds $C_t$ back to the PEVGs through the SEM and the PEVGs
play the jointly convex generalized Nash game of
Algorithm~\ref{algorithm1} for the allocation of energy within
$\sum_n x_n^t\leq C_t$.

Here,  $(\mathbf{x^1}^*, \mathbf{x^2}^*, ...,\mathbf{x^T}^*)$ and
$(p_1^*, p_2^*,...,p_T^*)$ constitute the solution of the discrete
time Stackelberg game under a feedback information structure with
the corresponding state information $(C_1, C_2,...,C_T)$, if
$\mathbf{x}{^t}^*$ comprises the solution of the GNE among the
followers for price $p_t^*$ at each $t=1,2,...,T$~\cite{PLM2004}.
The solution will be team optimal if the solution at the GNE is
optimal, that is if the GNE is a VE~\cite{FF07}, for the sub game in
each $t=1,2,...,T$~\cite{PLM2004}. Now, given that the Stackelberg
game described in Section~\ref{NGSE} constitutes a sub-game in each
time interval $t$ of the feedback Stackelberg game, the sub-game
will reach its optimal solution (as shown in
Section~\ref{existenceUniqueness}) within the constraint
$\sum_nx_n^t\leq C_t$ at each $t=1,2,...,T$. Therefore, the solution
of the discrete time feedback Stackelberg game possesses a team
optimal solution.
% -------------------------------------------

\section{Numerical analysis}\label{NumAnalysis}
\begin{figure}[t]
\centering
\includegraphics[width=8cm]{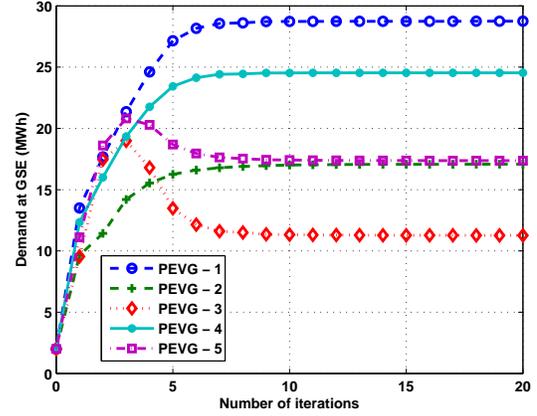}
\caption{Convergence of the demand of each PEVG to the GSE.}
\label{fig_Demand_at_GSNE}
\end{figure}
% -------------------------------------------
\begin{figure}[t]
\centering
\includegraphics[width=8cm]{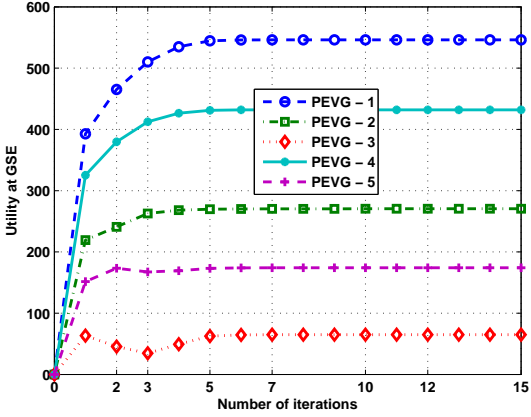}
\caption{Convergence of the utility of each PEVG to the GSE.}
\label{fig_Utility_at_GSNE}
\end{figure}
% --------------------------------------------
For our simulations, we consider a number of PEVGs that are
connected to the grid during peak hours. Here, a single PEVG entity
represents $1000$ vehicles at a specific location \cite{NR11}, where
each single vehicle is assumed to require $22$~kWh for every $100$
miles \cite{CS09,Tesla}. The maximum battery capacity of any single
vehicle is chosen between $150$ and $300$ miles. Hence, the maximum
capacity $b_n$ of any PEVG $n$ ranges between $35$~MWh and $65$~MWh,
which is chosen randomly for each PEVG. The available energy at the
grid is chosen as $99$ MWh. Initially, the grid sets the selling
price to $17$ USD per MWh. {We note that the chosen parameters
correspond to a typical PEVG use case~\cite{NR11,CS09,Tesla}.
However, we duly highlight that these parameters can vary
considerably according to PEVG usage and type, and economic
conditions within a given city, or country.}

The satisfaction parameter $s_n$ is chosen randomly in the range of
[$1,2$]. The range of $s_n$ is chosen based on the assumption that
the PEVG with the lowest satisfaction parameter $s_n=1$ will get
full satisfaction from each unit of energy it will consume whereas
the PEVG with the highest parameter i.e., $s_n=2$, will reach the
same satisfaction level from consuming half the amount due to its
smaller limitations in terms of its initial battery state and travel
plan\footnote{{Although a PEVG may not be able to demand the amount
of energy equal to its total battery capacity due to the scarcity of
energy at peak hour, it can ask for an amount that must be satisfied
in order for its constituent PEVGs to reach their satisfactions.}}.
We do not consider the PEVGs with $s_n>2$ as they do not have an
immediate need for energy at the peak hour. All statistical results
are averaged over all possible random values of the PEVGs'
capacities using around $1000$ independent simulation runs.

In Fig.~\ref{fig_Demand_at_GSNE} and Fig.~\ref{fig_Utility_at_GSNE},
 we show the demand and the utility at the GSE for a network with ${N}=5$ PEVGs.
Here, we can see that
 a similar demand by different PEVGs does
not always lead to a similar utility for the PEVGs. For example,
although PEVGs $2$ and $5$ in Fig.~\ref{fig_Demand_at_GSNE} have
almost the same demand at the GSE, their utilities are different
from one another as shown in Fig.~\ref{fig_Utility_at_GSNE}. This is
due to their different battery capacities and satisfaction
parameters. From the utility in \eqref{utility_function}, we can see
that the maximum utility level of a PEVG varies significantly for
different values of $b_n$ and $s_n$. Therefore, with the same energy
consumption, different PEVGs may obtain a different utility (e.g.,
PEVGs~$2$ and $5$). Fig.~\ref{fig_Demand_at_GSNE} and
\ref{fig_Utility_at_GSNE} show that, after the $10^{\text{th}}$
iteration, all the PEVGs reach their maximum utilities, and, thus,
their demands converge to the GSE.

\begin{figure}[t]
\centering
\includegraphics[width=8cm]{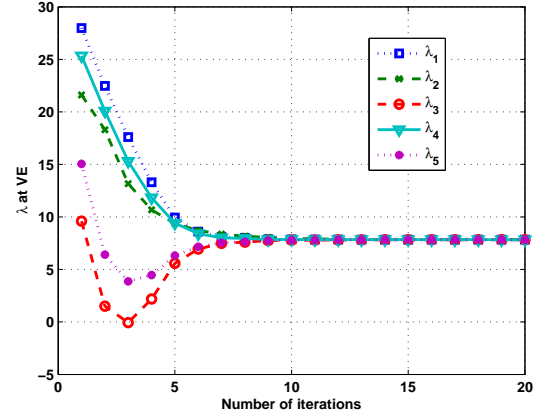}
\caption{Convergence of the values of  $\lambda_n,~\forall n$ to
$\lambda\geq 0$ as the solution of the GSG converges to GSE.}
\label{fig-lambdaGSNE}
\end{figure}

In Fig.~\ref{fig-lambdaGSNE}, we analyze the convergence speed of
the proposed algorithm by plotting the values of $\lambda_n$ as a
function of the number of iterations for $N=5$ PEVGs. First, recall
that the demands of all PEVGs converge to GSE for the optimal price
$p^*$ when all the PEVGs reach their VE and at this VE $\lambda_n =
\lambda\geq 0,~\forall n$. Fig.~\ref{fig-lambdaGSNE} shows that our
algorithm converges to the GSE after $9$ iterations (i.e., the PEVGs
reach their VE). Hence, as shown by Fig.~\ref{fig-lambdaGSNE}, the
convergence speed of our algorithm is reasonable.

% -------------------------------------------
% -------------------------------------------

\begin{figure}[t]
\centering
\includegraphics[width=8cm]{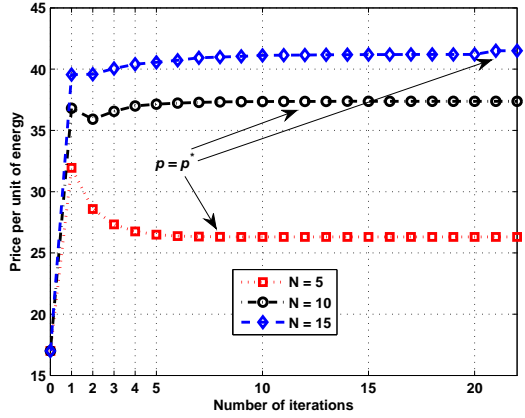}
\caption{Convergence of the price per unit of energy to the optimal
price for a network with different numbers of PEVGs.}
\label{figure-sp}
\end{figure}

In Fig.~\ref{figure-sp}, we show how the price set by the grid
converges to its optimal value as the strategies of the PEVGs
converge to the VE for networks of various sizes ($N=5,10$, and $15$
PEVGs)  with an initial  SG price set to $17$ USD per unit of
energy. Fig.~\ref{figure-sp} shows that the price converges to an
approximate optimal value within $5$ iterations. This is due to the
fact that the SG sets its price in response to the demand strategies
of the PEVGs. In fact, the PEVGs reach an approximate GSE within
five iterations, as shown for $5$ PEVGs in
Fig.~\ref{fig_Demand_at_GSNE}, and, then, they optimize their
demands within constraint \eqref{DemandCond}. The PEVGs eventually
reach the unique GSE that maximizes their utilities under this
constraint. Hence, the price $p$ converges quickly to its optimal
value $p^*$.
 % -------------------------------------------
% -------------------------------------------
Furthermore, Fig.~\ref{figure-sp} shows that the variation of the
grid's price is more noticeable when fewer PEVGs exist. This is due
to the fact that, for a fixed grid capacity, as the number of PEVGs
increases, there are fewer possibilities of variations in the
demands due to \eqref{DemandCond}.

\begin{figure}[t]
\centering
\includegraphics[width=8cm]{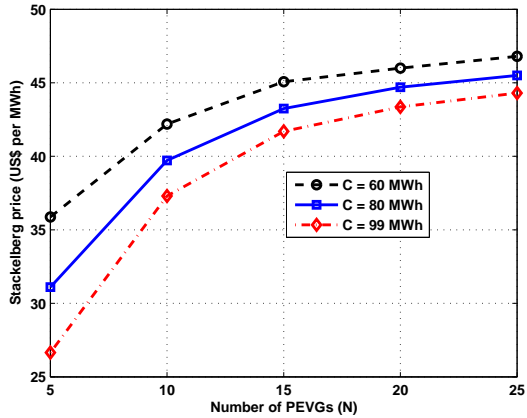}
\caption{Effect of increasing the number of PEVGs ${N}$ in the
network  and the grid energy $C$ on the average Stackelberg price.}
\label{fig-effect_NO_PEV_SP}
\end{figure}

In Fig.~\ref{fig-effect_NO_PEV_SP}, we show the effect of the number
of PEVGs on the optimal price choice of the grid. To do so, we
increase the number of PEVGs in the network for different grid
capacities $C = 60, 80$ and $90$ MWh.
Fig.~\ref{fig-effect_NO_PEV_SP} shows that the average optimal price
increases with the number of PEVGs in the network because of the
increasing energy demand on the SG's limited resources.
% -------------------------------------------
% -------------------------------------------
In contrast, increasing the grid's capacity leads to a decrease in
the optimal price $p^*$. This is due to the fact that, as the total
available capacity of the grid increases, the grid has more energy
to sell and, thus, it can decrease its price while maintaining
desirable revenues. Moreover, for a fixed number of PEVGs, as the
available energy at the grid increases, the SG reduces its optimal
price\footnote{We assume that the initial price is chosen based on
statistical estimation such as in~\cite{RW2012}, so that the grid
can maintain a revenue with optimized price as the grid capacity
increases.} to encourage the PEVGs to demand more energy.

\begin{figure}[t]
\centering
\includegraphics[width=8cm]{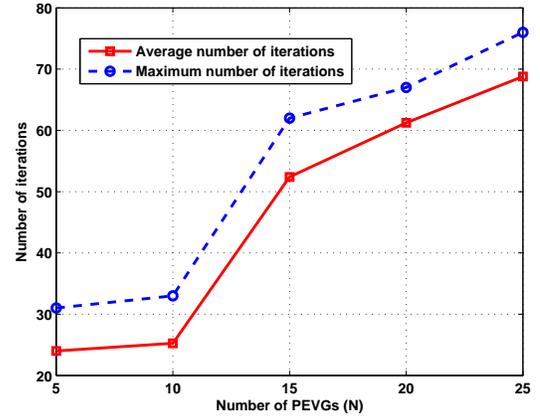}
\caption{Effect of increasing the number of PEVGs on the average and
maximum number of iterations to reach the GSE.}
\label{fig-no-of-iteration}
\end{figure}

Fig.~\ref{fig-no-of-iteration} shows the average and maximum number
of iterations needed to reach the GSE of the proposed game with
respect to the number of PEVGs.
% -------------------------------------------
% --------------------------------------------
In Fig.~\ref{fig-no-of-iteration}, we can see that, whenever the
number of PEVGs in the network increases, for the same amount of
capacity from the grid, the PEVGs require more iterations to reach
their optimal demands. For example, when the number of PEVGs in the
network increases from $15$ to $25$, the average number of
iterations needed to reach the GSE increases from $52$ to $79$.
Similar behavior is also seen for the maximum number of iterations.

{In Fig.~\ref{fig-utility_ED_GT}, we compare the results of the
proposed scheme with a particle swarm optimization (PSO)
\cite{WM2011} and an equal distribution (ED) scheme~\cite{CC04}. In
a PSO scheme, a group of random solutions\footnote{In this case the
energy demand of a PEVG.} (i.e., particles) are scattered over the
search space, and the particles converge to a near optimal solution
after a number of iterations. For the PSO algorithm, the particle
size is considered to be $40$. The parameters are updated in such a
way that the constraint in \eqref{DemandCond} is satisfied. For the
ED scheme, the available grid energy is distributed equally among
the connected PEVGs in the network. That is, if the available energy
at the grid is $C$ MWh and there are $N$ PEVGs connected to the
grid, then each PEVG receives an allocation of $\frac{C}{{N}}$ MWh
of energy from the grid as long as this allocation does not exceed
the maximum that a PEVG can be charged.}
% -------------------------------------------

{Fig.~\ref{fig-utility_ED_GT} shows that the average utility
achieved by the proposed scheme is better for most of the PEVGs in
the network, except PEVGs $9$ and $10$, when compared to the PSO
scheme. This is due to the fact that the PSO scheme optimizes the
energy for each PEVG according to the better particle position in
the available energy space. This may lead to a better utility for
PEVGs $9$ and $10$ at the expense of much lower utilities for the
rest of the PEVGs. However, the proposed scheme allocates the energy
among the PEVGs so that the socially optimal solution is achieved.
Therefore, most of the PEVGs in the network achieve improved
utilities compared with the PSO scheme. From
Fig.~\ref{fig-utility_ED_GT}, the proposed scheme has a total
utility, on the average, $1.3$ times the utility achieved by the PSO
scheme. Moreover, the proposed scheme has, on the average, twice the
utility achieved by the ED scheme which is a significant
improvement.

% -------------------------------------------
\begin{figure}[t]
\centering
\includegraphics[width=8cm]{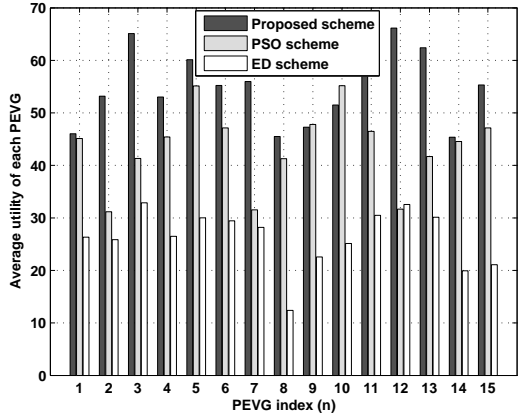}
\caption{Comparison of the average utilities of the PEVGs at the GSE
for the proposed scheme, the PSO scheme and ED scheme.}
\label{fig-utility_ED_GT}
\end{figure}
% --------------------------------------------

{Fig.~\ref{fig-demandvsPEV} shows the average demand per PEVG as the
number of PEVGs varies. In Fig.~\ref{fig-demandvsPEV}, we can see
that the average demand per PEVG decreases as the number of PEVGs
increases. This is a direct result of \eqref{DemandCond} and of the
fact that the optimal price set by the grid increases as the number
of PEVGs increases. From Fig.~\ref{fig-demandvsPEV}, we can see that
average PEVG demand for our scheme, the PSO scheme, and the ED case
decreases as the number of PEVGs increases.}
{Fig.~\ref{fig-demandvsPEV} shows that the energy demand for the
proposed scheme is lower than that for both the PSO and ED scheme
for all $N$. This can be interpreted by the fact that, in our
proposed scheme, each PEVG demands a required amount of energy
within constraint \eqref{DemandCond} based on its satisfaction
parameter and available grid energy. Therefore, each PEVG requests a
\emph{socially optimal} amount of energy from the grid rather than a
predefined amount (as in the ED scheme), or near optimal amount
according to a random search space (as in the PSO scheme). Clearly,
Fig.~\ref{fig-demandvsPEV} shows that the proposed scheme leads to
better energy utilization than either the PSO or ED scheme.}
% -------------------------------------------
\begin{figure}[t]
\centering
\includegraphics[width=8cm]{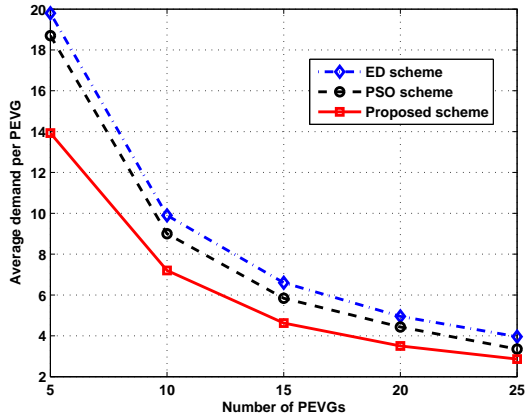}
\caption{Effect of the number of PEVGs in the network on the average
demand of a PEVG.} \label{fig-demandvsPEV}
\end{figure}
% -------------------------------------------

In Fig.~\ref{fig-utilityvsPEV}, we show the average utility achieved
by all three schemes as a function of the number of PEVGs. In this
figure, we can see that the average utility per PEVG decreases as
$N$ increases for all three schemes. This is due to the fact that
the benefit extracted by each PEVG decreases as more PEVGs share the
fixed available energy from the grid. However, importantly,
Fig.~\ref{fig-utilityvsPEV} shows that the proposed scheme has a
performance advantage at all network sizes which is, on the average,
$1.6$ times that of the PSO scheme. When compared to the ED scheme,
the proposed scheme shows significant improvements for all network
sizes, reaching an improvement of up to $3.5$ times over the ED
scheme for $N=25$ PEVGs.
% -------------------------------------------
\begin{figure}[t]
\centering
\includegraphics[width=8cm]{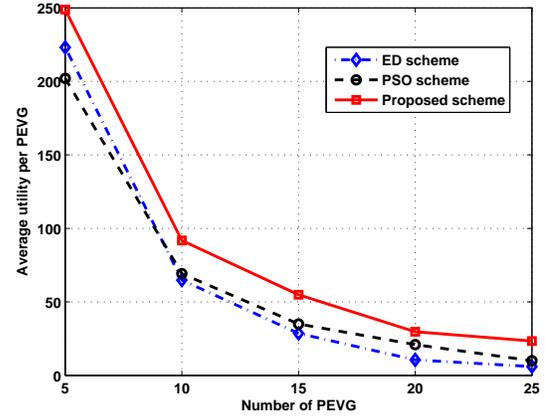}
\caption{Effect of the number of PEVGs in the network on the average
utility per PEVG.} \label{fig-utilityvsPEV}
\end{figure}
% -------------------------------------------

{In Fig.~\ref{fig-dynamic-allocation}, we assess the average demand
of energy per PEVG in a time-varying environment. Here, the state
transition of the variable $C_t$ in \eqref{state-equation} is
modeled as an independent stochastic process~\cite{KB2011}, due to
the random changes in traffic conditions and grid energy from one
time instant to another~\cite{FRAD2010}. The state of available
energy at the grid at any time $t$ is assumed to be a uniformly
distributed random variable in the range [$0.5,1.5$] times the
average available energy~\cite{FRAD2010}. The capacity of the PEVGs
$b_n^t~\forall n\in\mathcal{N}~\text{at}~t=1,2,.....,T$, is assumed
to be a uniformly distributed random variable in the range
[$0.5,1.5$] times the average battery capacity ~\cite{FRAD2010}.
Assuming that the traffic conditions in any PEVG changes every $30$
minutes, we have run independent simulations for allocating energy
to the PEVGs for eight time slots in peak hours (from 12 pm to 4
pm). The average available energy in the grid is $66$ MWh (assuming
a range from $33$ to $99$~MWh), and the variation of energy across
time slots is modeled by a uniformly distributed random variable
between $0.5$ and $1.5$ of this amount~\cite{FRAD2010}. Similarly,
the random variation in PEVGs' capacity at different time slots is
captured assuming a maximum of $55$ MWh (for $1000$ PEVs in a PEVG)
and a minimum of $9.9$ MWh (for $300$ PEVs)~\cite{EC08}.

In Fig.~\ref{fig-dynamic-allocation} it is shown that the demand of
any PEVG varies across time slots due to variation in both
satisfaction parameters of the PEVGs and the available energy.
However, the minimum energy demand by any PEVG is always well above
its minimum battery requirement. For example, the minimum demand by
PEVG $4$ is $5.5$ MWh in time slot $1$, which exceeds the minimum
battery requirement of a mid-size car, assuming $1000$ PEVs are in
that PEVG~\cite{ANRV2007}.
% -------------------------------------------
\begin{figure}[t]
\centering
\includegraphics[width=8cm]{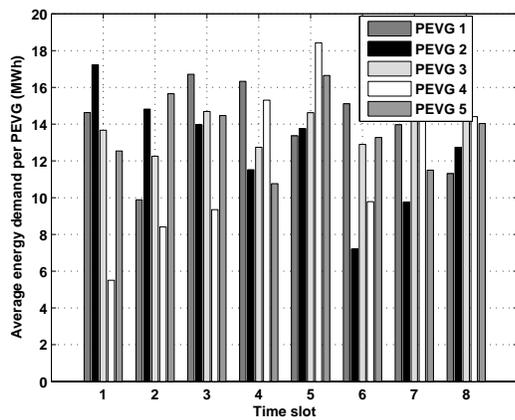}
\caption{Average demands of the PEVGs at the GSE for the proposed
scheme in a dynamic environment.} \label{fig-dynamic-allocation}
\end{figure}

The change of utility per PEVG in a time-varying environment is
shown in Fig.~\ref{fig-utility-dynamic} for the proposed, PSO and ED
schemes. Fig.~\ref{fig-utility-dynamic} shows that the team optimal
solution achieved by the proposed scheme leads to an improved
average utility for the PEVG when compared to the PSO and ED
schemes. However, the improvement varies across time slots due to
variations in available energy and the number of PEVs in the smart
grid network. As shown in Fig.~\ref{fig-utility-dynamic}, the
average utility achieved per PEVG by our proposed scheme is, on the
average, $1.6$ times the utility achieved by the PSO scheme and
$3.8$ times that achieved by the ED scheme.
% -------------------------------------------
\begin{figure}[t]
\centering
\includegraphics[width=8cm]{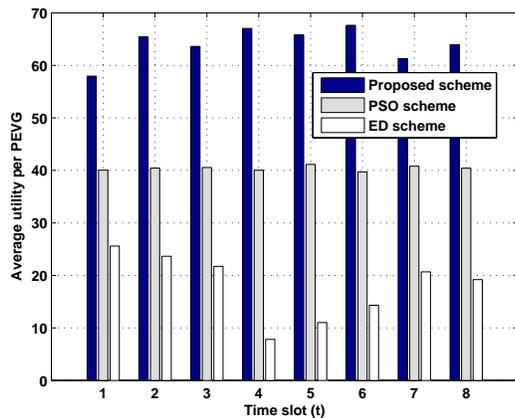}
\caption{Comparison of the average utility per PEVG in the dynamic
case.} \label{fig-utility-dynamic}
\end{figure}
% -------------------------------------------
\section{Conclusions}\label{conclusion}
In this paper, we have formulated a noncooperatibe Stackelberg game
to study the problem of energy trading between an SG and a number of
PEV groups. In this game, the SG chooses its price to maximize its
revenue whereas the PEVGs strategically choose the amounts of energy
they wish to buy from the grid so as to optimize a tradeoff between
the benefit of battery charging and associated costs. We have
studied the properties of this solution and we have shown that this
game admits a socially optimal generalized Stackelberg equilibrium.
We have also extended and analyzed the proposed game in a
time-varying environment. To reach the equilibrium of the game, we
have proposed a novel algorithm that can be adopted by the PEVGs, in
a distributed manner. Simulation results have shown that the
proposed approach yields improved performance gains, in terms of the
average utility per PEVG, compared to a particle swarm optimization
and an equal distribution scheme. Several future extensions can be
foreseen for this work, such as handling rapidly changing dynamics
or determining the optimal deployment of charging stations depending
on a variety of factors such as the parking duration, battery state,
and the location of the vehicles.

\vspace{-30pt}
\begin{IEEEbiography}{Wayes Tushar}
received his B.Sc. degree in Electrical and Electronic Engineering from Bangladesh University of Engineering and Technology  in 2007.  He is currently a Ph.D. Candidate at the Research School of Engineering in Australian National University (ANU). From May 2011 to June 2011 he was a visiting student research collaborator at the School of Engineering and Applied Science in Princeton University. He received the best student paper award in the Australian Communications Theory Workshop in 2012. His research interests include smart grid systems, game theory and signal processing.
\end{IEEEbiography}
\vspace{-20pt}
\begin{IEEEbiography}{Walid Saad}
received his B.E. from the Lebanese University in 2004, his M.E. in Computer and Communications Engineering from the American University of Beirut in 2007, and his Ph.D degree from the University of Oslo in 2010. From August 2008 till July 2009 he was a visiting scholar at the Coordinated Science Laboratory at the University of Illinois at Urbana Champaign. From January 2011 till July 2011, he was a Postdoctoral Research Associate at the Electrical Engineering Department at Princeton University.

Currently, he is an Assistant Professor at the Electrical and Computer Engineering Department at the University of Miami. His research interests span the areas of game theory, wireless networks, and the smart grid. He was the first author of the papers that received the Best Paper Award at the 7th International Symposium on Modeling and Optimization in Mobile, Ad Hoc and Wireless Networks (WiOpt), in June 2009 and at the 5th International Conference on Internet Monitoring and Protection (ICIMP) in May 2010. He is a co-author of the paper that won the best paper award at the IEEE Wireless Communications and Networking Conference (WCNC) in 2012.
\end{IEEEbiography}
\vspace{-20pt}
% Prof. H. Vincent Poor
\begin{IEEEbiography}{H. Vincent Poor}
(S'72, M'77, SM'82, F'87) received the Ph.D. degree in EECS from
Princeton University in 1977.  From 1977 until 1990, he was on the
faculty of the University of Illinois at Urbana-Champaign. Since
1990 he has been on the faculty at Princeton, where he is the
Michael Henry Strater University Professor of Electrical Engineering
and Dean of the School of Engineering and Applied Science. Dr.
Poor's research interests are in the areas of stochastic analysis,
statistical signal processing, and information theory, and their
applications in wireless networks and related fields such as social
networks and smart grid. Among his publications in these areas are
the recent books \emph{Classical, Semi-classical and Quantum Noise}
(Springer, 2012) and \emph{Smart Grid Communications and Networking}
(Cambridge University Press, 2012).

Dr. Poor is a member of the National Academy of Engineering and the
National Academy of Sciences, a Fellow of the American Academy of
Arts and Sciences, and an International Fellow of the Royal Academy
of Engineering (U.K.). He is also a Fellow of the Institute of
Mathematical Statistics, the Acoustical Society of America, and
other organizations.  In 1990, he served as President of the IEEE
Information Theory Society, and in 2004-07 he served as the
Editor-in-Chief of the \emph{IEEE Transactions on Information
Theory}. He received a Guggenheim Fellowship in 2002 and the IEEE
Education Medal in 2005. Recent recognition of his work includes the
2010 IET Ambrose Fleming Medal, the 2011 IEEE Eric E. Sumner Award,
the 2011 Society Award of the IEEE Signal Processing Society, and
honorary doctorates from Aalborg University, the Hong Kong
University of Science and Technology and the University of
Edinburgh.
\end{IEEEbiography}
\vspace{-15pt}
\begin{IEEEbiography}{David Smith}
is a Senior Researcher at National ICT Australia (NICTA) and is an
adjunct Fellow with the Australian National University (ANU), and
has been with NICTA and the ANU since 2004. He received the B.E.
degree in Electrical Engineering from the University of N.S.W.
Australia in 1997, and while studying toward this degree he was on a
CO-OP scholarship. He obtained an M.E. (research) degree in 2001 and
a Ph.D. in 2004 both from the University of Technology, Sydney
(UTS), and both in Telecommunications Engineering. His research
interests are in technology and systems for wireless body area
networks; game theory for distributed networks; mesh networks; radio
propagation and electromagnetic modeling; MIMO wireless systems;
coherent and non-coherent space-time coding; and antenna design,
including the design of smart antennas. He also has research
interest in optimization for smart grid. He has also had a variety
of industry experience in electrical engineering; telecommunications
planning; radio frequency, optoelectronic and electronic
communications design and integration. He has published numerous
technical refereed papers and made various contributions to IEEE
standardization activity; and has received a best paper award as
first author at 1st  International Symposium on Applied Sciences on
Biomedical and Communication Technologies, 2008, (ISABEL '08); and
has two other conference best paper awards, one as first author and
another as co-author.\end{IEEEbiography}
\end{document}